\documentclass[a4paper,
              ]{jacow}
\makeatletter%
	\ifboolexpr{bool{xetex}}
	 {\renewcommand{\Gin@extensions}{.pdf,%
	                    .png,.jpg,.bmp,.pict,.tif,.psd,.mac,.sga,.tga,.gif,%
	                    .eps,.ps,%
	                    }}{}
\makeatother

\ifboolexpr{bool{xetex} or bool{luatex}} 
 {}                                      
 {\usepackage[utf8]{inputenc}}           

\usepackage[USenglish]{babel}			 

\usepackage[final]{pdfpages}
\usepackage{multirow}
\usepackage{ragged2e}

\ifboolexpr{bool{jacowbiblatex}}%
 {%
  \addbibresource{jacow-test.bib}
  \addbibresource{biblatex-examples.bib}
 }{}
\begin{document}

\title{Potential and Issues for Future Accelerators and Ultimate Colliders}

\author{S. J. Brooks\thanks{sbrooks@bnl.gov}, Brookhaven National Laboratory, Upton, Long Island, New York}
	
\maketitle

\begin{abstract}
Particle colliders have been remarkably successful tools in particle and nuclear physics.  What are the future trends and limitations of accelerators as they currently exist, and are there possible alternative approaches?  What would the ultimate collider look like?  This talk examines some challenges and possible solutions.  Accelerating a single particle rather than a thermal distribution may allow exploration of more controlled interactions without background.  Also, cost drivers are possibly the most important limiting factor for large accelerators in the foreseeable future so emerging technologies to reduce cost are highlighted.
\end{abstract}

\section{Motivation: Energy Frontier}
This talk is developed from a previous one the author gave at the F3iA 2016 meeting \cite{F3iA}, whose goal was to brainstorm ideas for accelerators 50--100 years from now.  It also cites a few ideas mentioned at that meeting.  For a fuller view of others' perspectives, readers are encouraged to view the slides on that meeting's website.

The goal of many currently-proposed energy frontier accelerators is to investigate particles postulated to exist at the few-TeV mass scale.  There is, however, the possibility that no such particles exist and the next energy scale for new physics is substantially higher.  In that case, what needs to be different about our approach in designing accelerators?  In the following, an extreme example is examined together with its implications for accelerator design.

\section{Towards the Planck Energy}
Without making any assumptions beyond the standard model, a situation that would guarantee access to new physics would be if an accelerator probed the Planck energy scale of $E_\mathrm{Pl} = \sqrt{\hbar c^5/G} = 1.22\times10^{16}$\,TeV = 1.96\,GJ.  The magnitude of this energy alone is not the problem, since the LHC in 2017 stored two beams of $3.1\times10^{14}$ protons \cite{LHCstats} at an energy of 6.5\,TeV, which totals 645\,MJ = 0.33$E_\mathrm{Pl}$.  The difficulty is that the energy is in too many particles.

One limiting case is to consider accelerating a single particle described by a wavefunction $\psi(\mathbf x)$.  This function would be deflected through magnets and accelerating structures in much the same way as a beam, only deviating when diffractive effects at the de Broglie wavelength $\lambda=2\pi\hbar/|\mathbf p|$ dominate.  The particles in our tracking codes would be replaced by wavepackets \cite{wavepacketnote}.  The uncertainty principle $\sigma_x\sigma_p\ge\hbar/2$ applies to this one-particle `beam', meaning that in more familiar units, it has a minimum normalised emittance of $\epsilon_{N,rms}\ge\hbar/2mc$ in each phase space plane.

The construction of an accelerator that reaches this minimum quantum emittance requires preparation of particles in their ground state, requiring techniques from atomic, quantum and ultra-cold physics.  These areas are not familiar to the average accelerator physicist (and that includes the author!)  They are, however, a useful pointer to where future collaborations should be.  An emerging nexus of these fields is in quantum computing, where the control of the wavefunction, extending to multi-particle `entangled' states, is paramount.  The alignment that would be required to collide such localised particles would bear more resemblance to work in metrology and gravitational wave detection.

\subsection{Accelerating Gradients}
The next parameter to examine is the length of the accelerator: a very high gradient would be required to reach the Planck energy in a practical length.  Examples of gradient levels are listed below:
\begin{itemize}
\item $\le10^8$\,V/m: conventional accelerating structures.
\item $10^9$--$10^{11}$\,V/m: laser-plasma structures.
\item $2\times10^{12}$\,V/m: proposed open plasma structures powered by coherent laser echelons generated by fibre lasers \cite{Pukhov}.
\item $10^{15}$\,V/m: maximum electric field produced at the ELI-NP laser \cite{ELI-NP}.
\item $1.32\times10^{18}$\,V/m $=m_e^2c^3/e\hbar$: Schwinger limit at which $e^-e^+$ pair production starts in strong electric fields and $\gamma\gamma$ scattering in light.
\item $2.65\times10^{18}$\,V/m $=2m_e^2c^3/e\hbar$: gradient corresponding to Caianiello's maximal acceleration \cite{Caianiello,Caianiello-Papini} of an electron.
\item $5\times10^{19}$\,V/m: gradient experienced by an $\alpha$ particle undergoing Rutherford back-scattering \cite{Rutherford} from a gold nucleus.
\item $1.13\times10^{23}$\,V/m $=2m_\mu^2c^3/e\hbar$: gradient corresponding to Caianiello's maximal acceleration of a muon.
\end{itemize}

Owing to the presence of several theoretical limits above $10^{18}$\,V/m and ongoing technological progress below that level, it seems reasonable to choose this as a limiting level of accelerating gradient in a far future facility.  Unfortunately, $E_\mathrm{Pl}/(e10^{18}\,\mathrm{V/m})$ = 12.2 million km.

\subsection{Shortcut to the Planck Scale: Black Hole}
A solution presents itself from the fact that the Schwarzschild radius of a black hole $r_s=2GM/c^2$ increases linearly with total mass-energy.  At the Planck energy, $r_s$ is of the same scale as the de Broglie wavelength that determines the minimal focal size of particles.  Thus, if $x$ times the Planck energy were focussed in total, particles with $x$ times longer wavelength could be used to form a black hole.  These particles only have of order $1/x$ the Planck energy each, so in the simplest case of bosons whose wavefunctions can overlap, of order $x^2$ particles would be required.  The black hole formed is predicted to undergo Hawking evaporation \cite{Hawking}, which is itself a test of Planck-scale physics as the hole shrinks to the Planck scale just before evaporating.

A Gaussian distribution of mass-energy with size $\sigma_x$, considered in flat space for simplification, would first form an event horizon at $r=2.14\sigma_x$, which contains 79.4\% of the total distribution.  If a convergence angle for particles is defined $\sigma_\theta=\sigma_p/p\simeq\sigma_p/(E/c)$ in the ultrarelativistic approximation, the lower bound on focus size is
\[ \sigma_x \ge \frac{\hbar}{2\sigma_p} = \frac{\hbar c}{2\sigma_\theta E}. \]
In the case of forming a black hole with $N$ particles of energy $E$, the equivalent mass within $2.14\sigma$ is $M=0.794NE/c^2$ and
\[ \sigma_x = \frac{r_s}{2.14} = \frac{2GM}{2.14c^2} = \frac{0.743GNE}{c^4}. \]
Setting this equal to the minimal focus size gives
\[ N =  \frac{c^4}{0.743GE} \frac{\hbar c}{2\sigma_\theta E} = \frac{0.673\hbar c^5}{G\sigma_\theta E^2} = \frac{0.673}{\sigma_\theta}\left(\frac{E_\mathrm{Pl}}{E}\right)^2. \]
For large numbers of particles, this total number can be evenly split between the two beams and gives, as an example, the parameters for the `bosons' column in Table \ref{params}.

\begin{table*}[hbt]
   \centering
   \caption{Black Hole Factory Parameter Table}
   \begin{tabular}{lcc}
       \toprule
       \textbf{Parameter} & \textbf{Bosons e.g. $\gamma$ (overlapping)} & \textbf{Fermions or non-overlapping bosons} \\
       \midrule
Energy&$10^{10}$\,TeV&$10^{12}$\,TeV\\
Length&10\,km&1000\,km (space)\\
Gradient&$10^{18}$\,V/m&$10^{18}$\,V/m\\
Particles per beam&$10^{12}$&$10^{12}$\\
Total energy per pulse&$3.22\times10^{15}$\,J = 893\,GW.h&$3.22\times10^{17}$\,J = 89.3 TW.h\\
Repetition period&14 days&14 days\\
Average power&2.66\,GW&266\,GW\\
$\sigma_x^* = \sigma_y^* = \sigma_z^*$&$1.97\times10^{-29}$\,m&$1.97\times10^{-27}$\,m (beam)\\
$\sigma_\theta^* = \sigma_E^*/E$&0.5\,rad = 50\%&0.5\,rad = 50\%\\
Black hole radius = $2.14\sigma_x^*$&$4.22\times10^{-29}$\,m&$4.22\times10^{-27}$\,m\\
Black hole mass&28.4 grams&2.84\,kg\\
Black hole lifetime \cite{Hawking-Page}&$1.10\times10^{-22}$\,s (evaporation)&$1.10\times10^{-16}$\,s\\
       \bottomrule
   \end{tabular}
   \label{params}
\end{table*}

In the case of fermions, $N$ particles will occupy $N$ times the 6D phase space volume, which can be split into a factor of $N^{1/3}$ emittance in each phase-space plane, so the beam obeys $\sigma_x\sigma_p\ge N^{1/3}\hbar/2$ and the minimum focus size is
\[ \sigma_x^\mathrm{Fermions} \ge \frac{\hbar c N^{1/3}}{2\sigma_\theta E}. \]
This carries through the previous calculation to give
\[ N = \frac{0.673N^{1/3}}{\sigma_\theta}\left(\frac{E_\mathrm{Pl}}{E}\right)^2 
\quad \Rightarrow \quad
N = \left(\frac{0.673}{\sigma_\theta}\right)^{3/2}\left(\frac{E_\mathrm{Pl}}{E}\right)^3, \]
which is a less favourable scaling than that of bosons and motivates the `fermions' column in Table \ref{params}.  
This non-overlapping configuration may also have to be used for photons if they scatter from each other as their intensities exceed the Schwinger limit on the way to the focus.  

\newcommand{\comment}[1]{}
\comment{
\begin{table*}[hbt]
   \centering
   \caption{Comparison of Parameters at 100\,km Length}
   \begin{tabular}{lcc}
       \toprule
       \textbf{Parameter} & \textbf{Bosons e.g. $\gamma$ (overlapping)} & \textbf{Fermions or non-overlapping bosons} \\
       \midrule
Energy&$10^{11}$\,TeV&$10^{11}$\,TeV\\
Length&100\,km&100\,km\\
Gradient&$10^{18}$\,V/m&$10^{18}$\,V/m\\
Particles per beam&$10^{10}$&$10^{15}$\\
Total energy per pulse&$3.22\times10^{14}$\,J = 89.3\,GW.h&$3.22\times10^{19}$\,J = 8.93 PW.h\\
Repetition period&14 days&14 days\\
Average power&266\,MW&26.6\,TW\\
$\sigma_x^* = \sigma_y^* = \sigma_z^*$&$1.97\times10^{-30}$\,m&$1.97\times10^{-25}$\,m (beam)\\
$\sigma_\theta^* = \sigma_E^*/E$&0.5\,rad = 50\%&0.5\,rad = 50\%\\
Black hole radius = $2.14\sigma_x^*$&$4.22\times10^{-30}$\,m&$4.22\times10^{-25}$\,m\\
Black hole mass&2.84 grams&284\,kg\\
Black hole lifetime \cite{Hawking-Page}&$1.10\times10^{-25}$\,s (evaporation)&$1.10\times10^{-10}$\,s\\
       \bottomrule
   \end{tabular}
   \label{params100km}
\end{table*}
}

The values in the Table \ref{params} describe a speculative facility to probe the Planck energy, which at least in the case of bringing bosons to a focus, does not present impossibly large size or energy demands.  The energy required to form each black hole would cost \$107M at current US electricity prices.  The gradient is below the Schwinger limit but the $10^{12}$ particles of each beam must be controlled to an unprecedented degree, maintaining the minimum quantum emittance.  The most obviously difficult parameter is the alignment of order $10^{-29}$\,m but in fact maintaining the small emittance is what requires radically-new methods, as shown in the next section.

\section{Reducing Scattering}
\begin{figure*}[!tbh]
    \centering
    \includegraphics*[width=\textwidth]{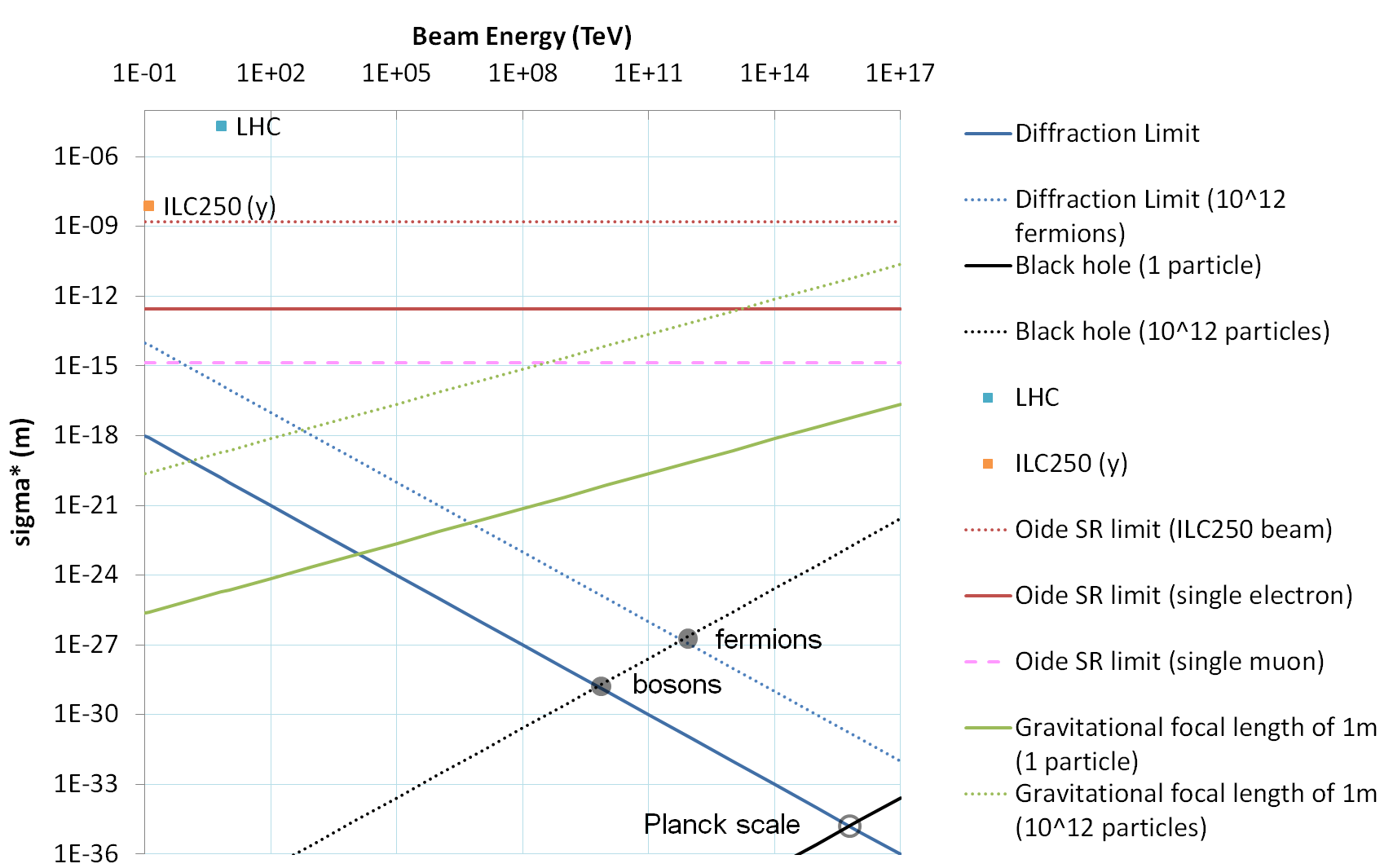}
    \caption{Highly logarithmic chart of particle energy vs. focal size.}
    \label{logchart}
\end{figure*}

Figure \ref{logchart} shows the space of $E$ vs. $\sigma_x$ ranging from current accelerators down to the Planck scale, with the two parameter sets from Table \ref{params} marked.  The diffraction limit and black hole formation form lower bounds on $\sigma_x$, however our current accelerators (particularly linear colliders) are nearing another limit found by Oide \cite{Oide}.  This comes about because the final bend before the interaction point causes particles in the beam to lose energy and scatter from synchrotron radiation; the resulting minimum focus size is nearly independent of everything apart from normalised beam emittance $\epsilon_N$.  The minimal emittance of $\hbar/2mc$ can be substituted into Oide's formula to find lower bounds for single electrons and muons, assuming the wavefunction itself behaves sufficiently like a beam in its scattering.  Proton beams will instead be limited by the actual size of the proton $8.75\times10^{-16}$\,m.

This limits linear colliders to a corridor where beam size at the IP cannot be reduced, which results in power demands that increase as $E$ or $E^2$ to achieve the same luminosity.  Being able to scale diagonally down the diagram in Figure \ref{logchart} would cancel this with improvements in $\sigma^*$.  The only way to do this is to find conditions under which the assumptions behind the Oide bound no longer apply, three of which are described in the following sections.

\subsection{(A) Bending happens at lower energy than focus}
\begin{figure}[!htb]
   \centering
   \includegraphics*[width=\linewidth]{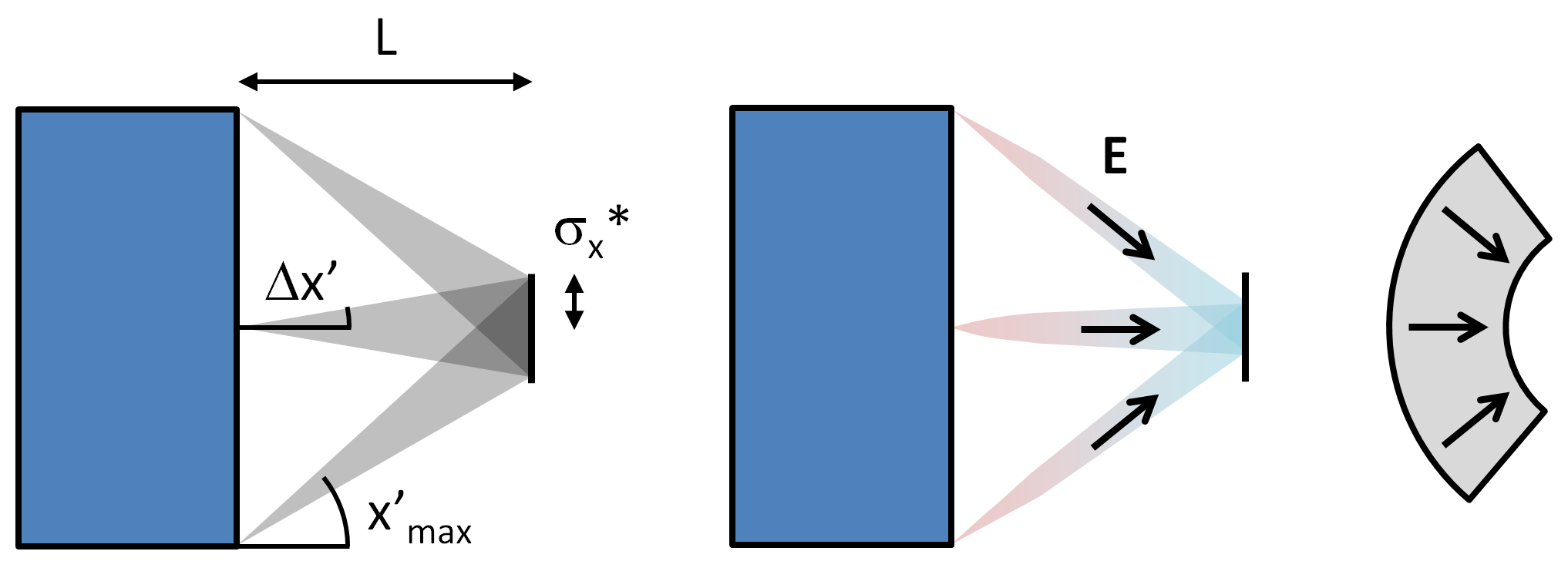}
   \caption{Schematic of a system that evades the Oide lower bound on focal size.  (Left) final focus magnet with beam having the minimal conventional focus size.  (Middle) Convergent accelerating \textbf{E}-field inserted parallel to the average beam propagation direction.  (Right) Pillbox cavity with spherical ends that produces a convergent \textbf{E}-field.}
    \label{typeA}
\end{figure}

If the lower bound on focal size is achieved and then acceleration is inserted between the final magnet and collision point, as shown in Figure \ref{typeA}, the focal size can be reduced further.

This is allowed because the derivation of the Oide bound assumes that bending is the final element before the collision point, not acceleration, meaning $\gamma$ and $\epsilon_\mathrm{geom}$ are equal in the magnet and collision point.  The \textbf{E}-field has to be parallel to the convergent direction of the beam because otherwise the acceleration will bend the particles by an angle comparable to the final magnet.  With a converging \textbf{E}-field, the only bending is proportional to the variation in angle at a given point in the beam (labelled $\Delta x'$), which can be made arbitrarily small relative to the full angle by increasing the distance of the final magnet from the minimal beam spot, which stays the same absolute size.

\subsection{(B) Quantum effects (coherence, entanglement)}
\begin{figure}[!htb]
   \centering
   \includegraphics*[width=\linewidth]{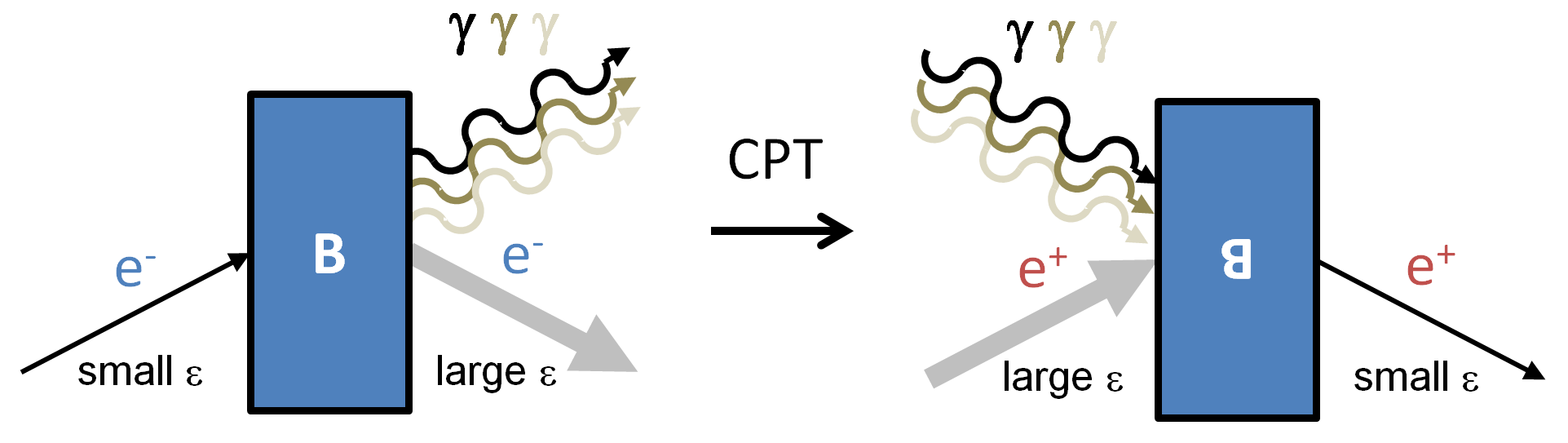}
   \caption{Applying the CPT symmetry to synchrotron radiation emission.}
    \label{typeBrev}
\end{figure}

In the derivation of emittance growth from synchrotron radiation, the emission of radiation is often treated as a statistical (i.e. incoherent) process.  However, quantum processes always have a time reverse as guaranteed by the CPT theorem.  Figure \ref{typeBrev} shows what this looks like for synchrotron emission.

\begin{figure}[!htb]
   \centering
   \includegraphics*[width=\linewidth]{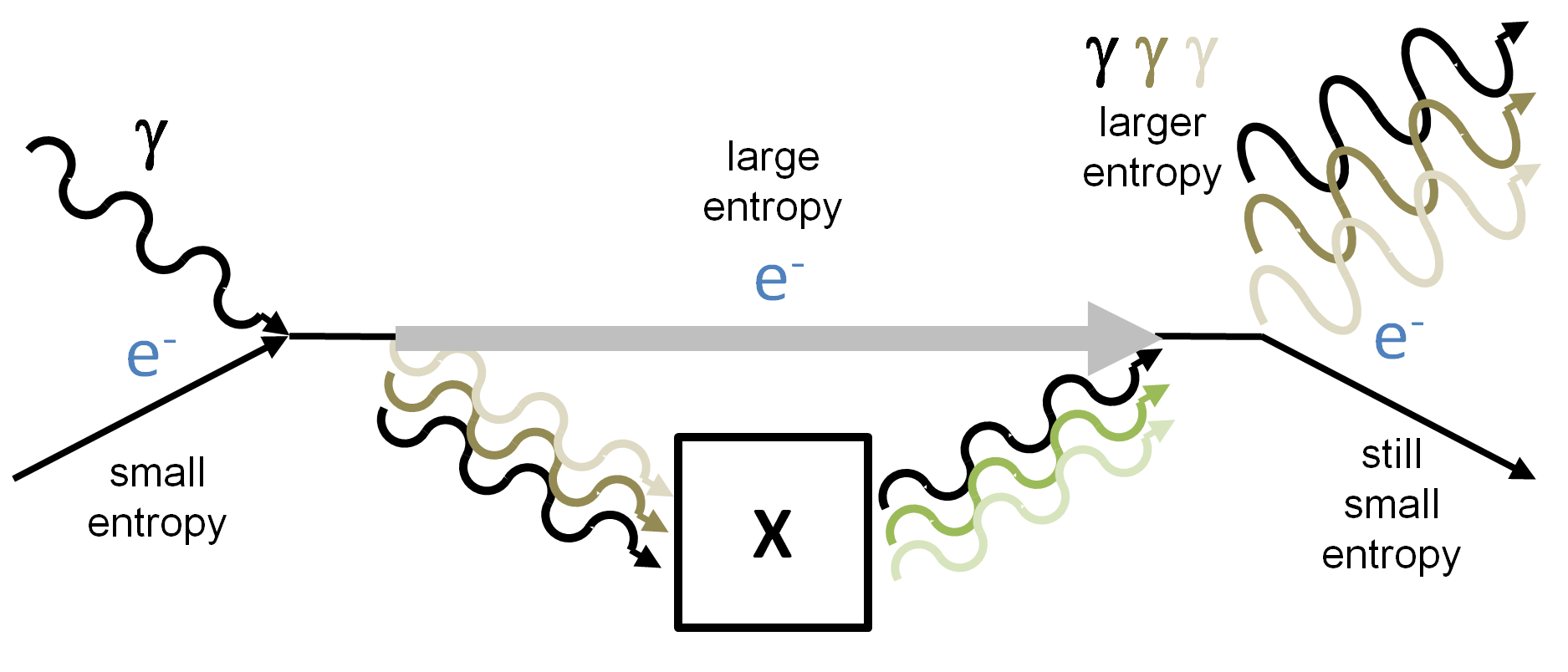}
   \caption{Schematic of a quantum cooling system.}
    \label{typeBcool}
\end{figure}

In Figure \ref{typeBrev}, a particular prepared photon state hits an with electron large emittance and is absorbed through the magnet, turning it into one with smaller emittance.  The difficulty is that this photon state is highly correlated with the electron, likely entangled with it, so would require special preparation.  The overall principle, on the other hand, is not so different from classical beam cooling methods, where information from the beam is extracted and then electromagnetic fields produced to reduce the beam's emittance.  The quantum analogy to cooling is shown in Figure \ref{typeBcool}, where the difficult task is to find the quantum operation X that results in transferring most of the entropy into the exhaust photon.

The CPT theorem also directly gives a low-energy state that will evolve into densities sufficient to create a black hole: start with the required high density focus and evolve it forward in time until it is absorbed by the (anti-matter) surroundings and converted into low-energy particles.  The CPT transformation of this result is the required (matter) initial state.  This is a vastly complex entangled state that is hard to make in practice, so procedures involving fewer particles like Figures \ref{typeBrev} and \ref{typeBcool} are more likely to be useful.

This construction resembles the idea of `Mössbauer acceleration' \cite{Muller}, where it is proposed that many entangled excited nuclei in a single crystal could emit all their energy in a single photon by exploiting the coherent interaction of the entire crystal as a recoiling body, as was observed for single nuclei in the traditional Mössbauer effect \cite{Mossbauer,Mossbauer-Craig}.  A Planck energy photon could be emitted from 30\,grams of $^{191}$Ir$^*$ in this manner.  One already-realised application of the Mössbauer effect is coherent control of the wavefunctions of individual gamma photons \cite{Vagizov}, which is the type of control required for single-particle acceleration.

Since particle accelerators generate both the beam and the accelerating fields, future experiments could entangle any combination of these with each other.  This would provide a large number of degrees of freedom unexplored by the incoherent thermal beam distributions used currently.

\subsection{(C) Non-electromagnetic focussing}
\begin{figure}[!htb]
   \centering
   \includegraphics*[width=\linewidth]{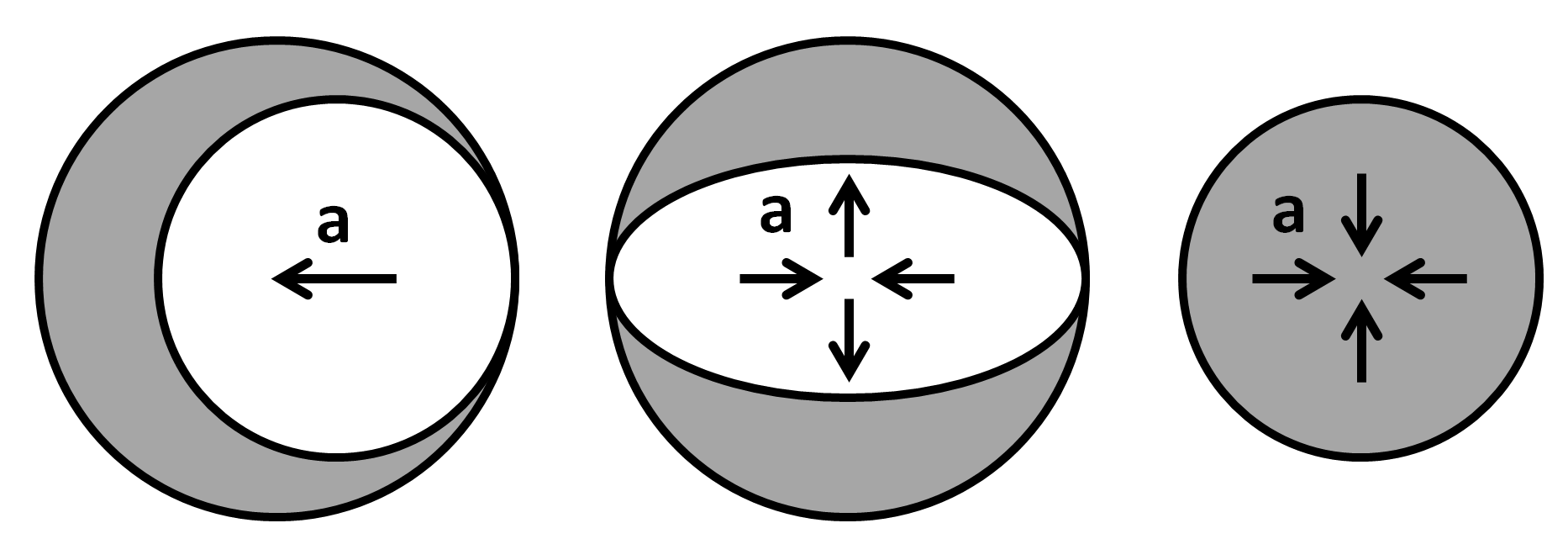}
   \caption{2D, completely-linear gravitational lenses, based on subtracting two K-V distributions of mass.  (Left) Dipole.  (Middle) Quadrupole.  (Right) Linear `monopole' focussing lens that would require the beams to interpenetrate.}
    \label{typeCgrav}
\end{figure}

An extreme way to avoid synchrotron radiation is to not use electromagnetic fields.  Gravity can focus trajectories of particles irrespective of charge and lenses can be constructed from dense matter as shown in Figure \ref{typeCgrav}.

The main difficulty is that in most situations gravity is far too weak to be useful.  Machines that produce black holes are the exception, as gravitational lenses bend by an angle $\theta\simeq2r_s/r$ where $r$ is the distance to the mass and $r_s$ is its Schwarzschild radius.  For example, if the focus angle $\sigma_\theta=0.5$\,rad in Table \ref{params} is difficult to achieve, an angle of 0.05\,rad would provide a 10$\times$ larger and weaker focus, which would bend adjacent particles by up to 0.2\,rad.  Several of these lenses could then produce the required angle for the final focus, at the expense of having more energy in flight.  The particles forming the lenses would not form black holes, so energy recovery may be possible.

\section{Nucleus-Level Alignment}
Collisions of spatially-confined and controlled particles present some interesting experiments at much easier scales than the $10^{-29}$\,m alignment required in the previous section.  Nanopositioner stages currently on the market have a resolution of  $5\times10^{-11}$\,m and the LIGO gravitational observatory has measured a length changes on the order of $10^{-18}$\,m, with its active suspension systems achieving $2\times10^{-13}$\,m \cite{LIGOsuspension}.  There may be technology available soon that can control position to the width of an atomic nucleus.  With a sufficiently cold target and focussed beam, reaction rates may be obtained in excess of what nuclear cross-sections would predict for random distributions.  

\begin{figure}[!htb]
   \centering
   \includegraphics*[width=0.7\linewidth]{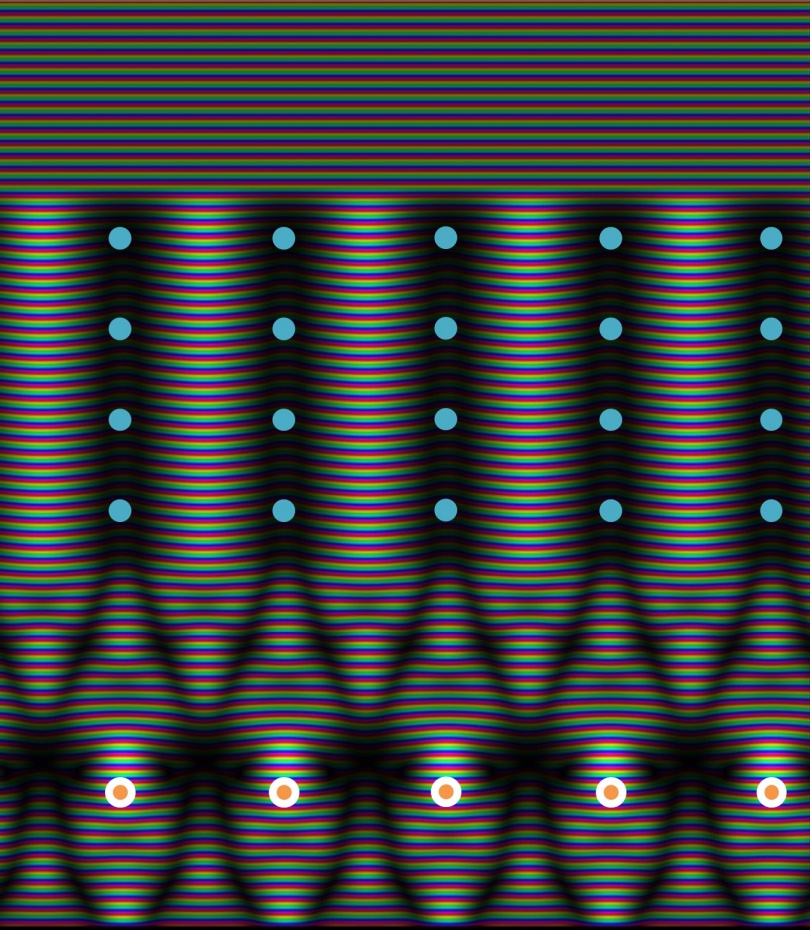}
   \caption{Alignment of an incoming plane-wave particle via crystal channelling (blue nuclei are the crystal), leading to amplitude fringes after leaving the crystal.  Target nuclei (orange) are be placed at the amplitude peaks.}
    \label{crystalfocus}
\end{figure}

Another route to positional enhancement of nuclear reaction rates is shown in Figure \ref{crystalfocus}, where an incoming de Broglie wave is confined to periodic potential wells in `crystal channelling' and then forms interference fringes on exiting the crystal.  Target nuclei could be placed at the amplitude peaks by nanotechnology thin film deposition methods.

\section{The Cheapness Frontier}
A complementary goal for future accelerators is to increase the capability of hardware per unit price, since today's accelerators are already some of the largest government-funded science projects in the world.  The following sections describe paradigms for doing this.

\subsection{Mass Production}
Custom parts can be more expensive than mass-produced parts by one or two orders of magnitude.  Research particle accelerators are custom systems, many of whose components are themselves custom, so can cost dramatically more than their constituent raw materials.

\begin{figure}[!htb]
   \centering
   \includegraphics*[width=\linewidth]{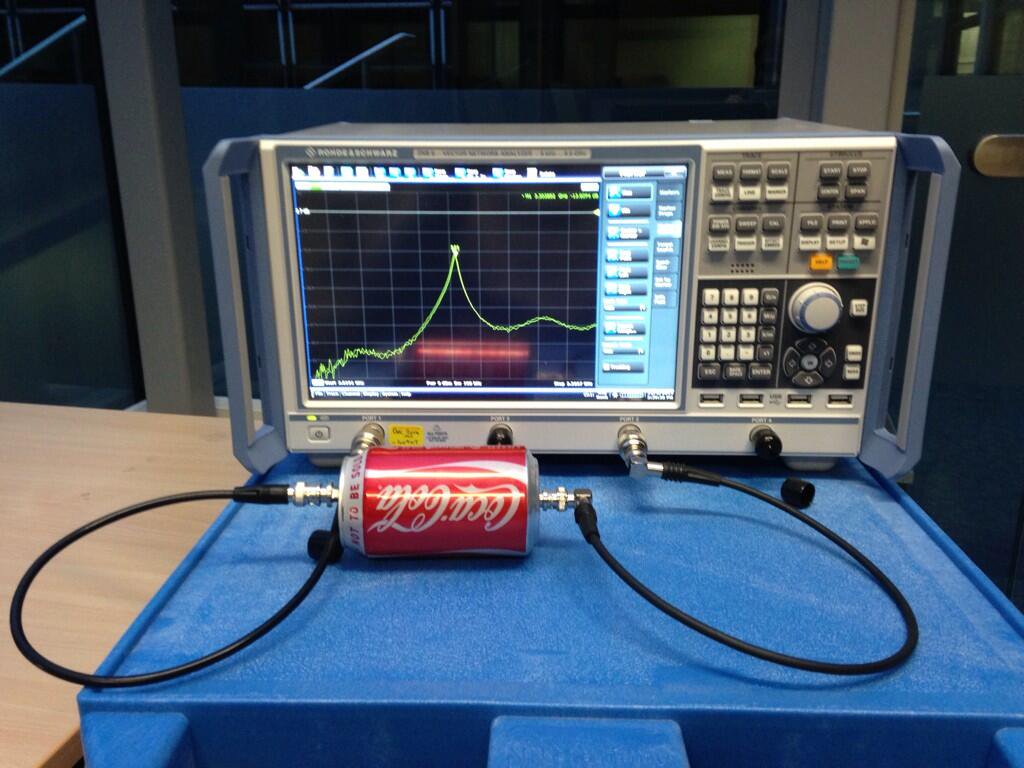}
   \caption{Mass-produced can showing resonance at $f$ = 3.3\,GHz, $Q$ = 50 \cite{Plostinar}.}
    \label{cokeRF}
\end{figure}

Occasionally a mass-produced component from another industry can be used.  Many metal containers are RF resonators as shown in Figure \ref{cokeRF} and beer barrels have been used as VHF resonators \cite{beer} with a quality factor $Q>9000$.  Computer displays are an example of a massively multi-channel voltage supply to millions of pixels, enabled by integrated circuits.

\subsection{Avoiding Tight Specification}
\begin{figure}[!htb]
   \centering
   \includegraphics*[width=\linewidth]{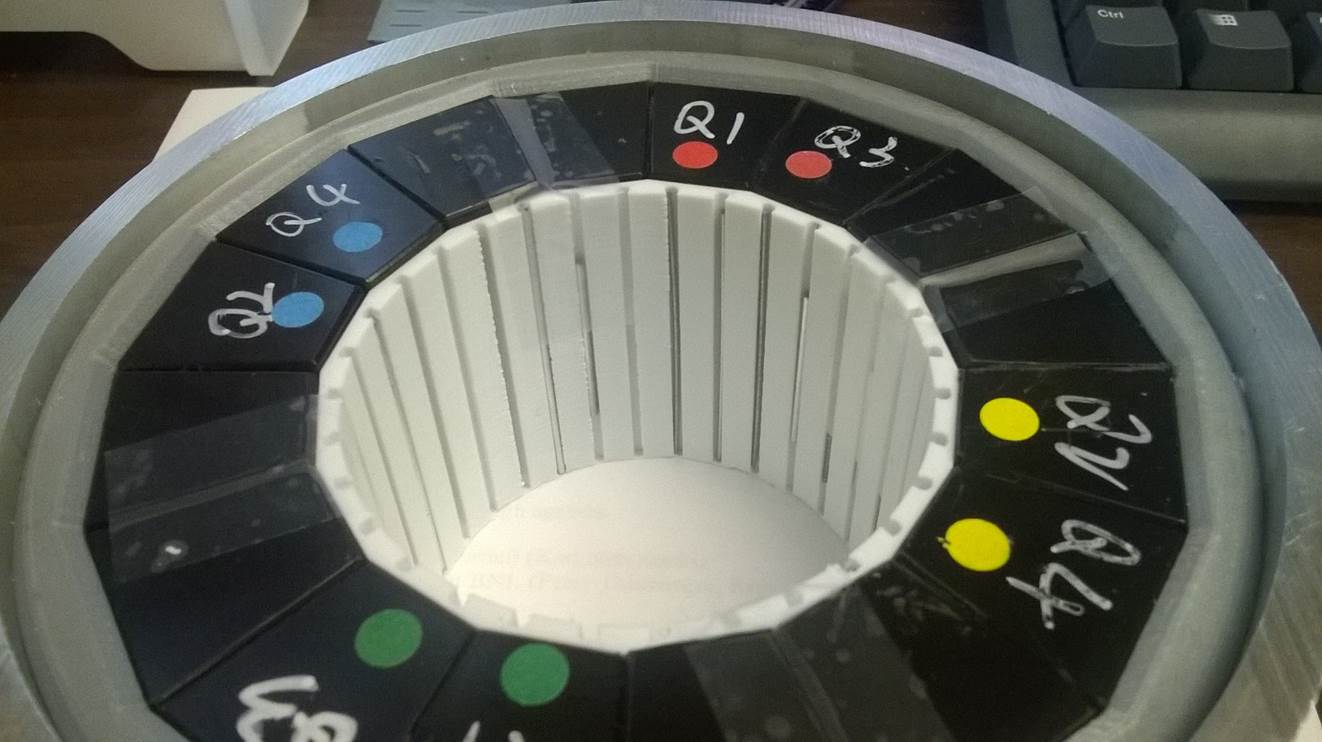}
   \caption{Halbach magnet using 3D printed mould and tuning insert containing iron wires.}
    \label{magnet}
\end{figure}

Having to manufacture a single component to very tight tolerances can inflate the cost by requiring special machining.  Often, precision can be obtained in a staged way by successively finer levels of adjustments or feedback.  Figure \ref{magnet} shows a permanent magnet described in \cite{magnets}, whose blocks are only accurate to 1\% but contains a tuning insert that can bring the field accuracy to $10^{-3}$ or $10^{-4}$.

\subsection{Automation}
If economic progress continues, people's time will become ever more costly relative to raw materials.  Robotics and 3D printing (again used in \cite{magnets}) can be used to highly automate the construction stage of accelerators, while automated tools incorporating optimisation and simulation can reduce manpower in the design stage.  Evolutionary algorithms can converge to a good accelerator design even starting from very general constraints \cite{thesis}.

\subsection{Re-use and Recycling}
Energy may be conserved by replacing electromagnets by superconducting or permanent magnets and copper RF by superconducting RF.  Beam energy recovery is an emerging solution for when a high power beam is not massively disrupted by its intended use.  Fixed-field alternating gradient lines allow multiple beam energies to be transmitted through a single line, even simultaneously, reducing the amount of hardware for recirculations of different energies.

\begin{figure}[!htb]
   \centering
   \includegraphics*[width=\linewidth]{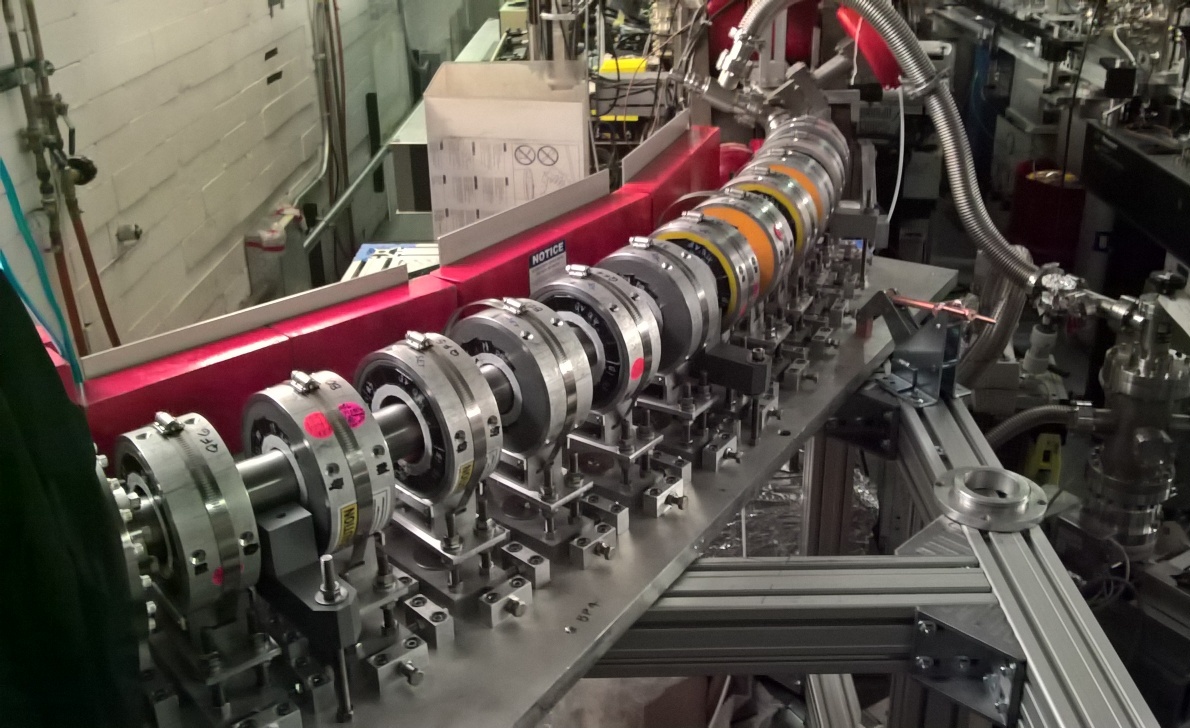}
   \caption{Fixed-field 40$^\circ$ arc with a 3.8$\times$ energy range (18--70\,MeV electrons), using permanent magnets.  Requires no power to operate.}
    \label{atf}
\end{figure}

Figure \ref{atf} shows a fixed-field test line that also uses permanent magnets \cite{ATF}.  The CBETA project \cite{CBETA,here} uses those technologies in the arcs, together with a 4-pass superconducting energy-recovery linac that aims to give 6\,MW of circulating beam with only 45\,kW of RF amplifiers.

\end{document}